\documentclass[a4paper,11pt]{article}

\usepackage{jheppub} 

\usepackage{tikz-cd}
\usepackage{soul}

\title{\boldmath Massless $AdS_2$ scattering and Bethe Ansatz}

 \author{A. Fontanella}
 \author{and A. Torrielli}
 \affiliation{Department of Mathematics, 
 \\University of Surrey,
 \\ Guildford, GU2 7XH, UK}

\emailAdd{a.fontanella@surrey.ac.uk}
\emailAdd{a.torrielli@surrey.ac.uk}

\abstract{We first analyse the integrable scattering theory describing the massless excitations of $AdS_2 \times S^2 \times T^6$ superstrings in the relativistic limit. The matrix part of the S-matrix is obtained in the BMN limit from the conjectured exact expression, and compared to known S-matrices with ${\cal{N}}=1$ supersymmetry in $1+1$ dimensions. A dressing factor, yet unknown for the complete theory, is here constructed based on relativistic crossing symmetry. We derive a Bethe-ansatz condition by employing a transfer-matrix technique based on the so-called {\it free-fermion condition}. This is known to overcome the problem of lack of a reference state. We then generalise the method to the massless {\it non-relativistic case}, and compare the resulting Bethe-ansatz condition with a simple massless limit of the one conjectured by Sorokin, Tseytlin, Wulff and Zarembo.}


\arxivnumber{1706.02634}

\dedicated{Dedicated to the memory of Professor Mario Tonin.}

\begin{document}

\maketitle
\flushbottom

\section{Introduction}
The remarkable success the integrability program has had in solving string theory on $AdS_5 \times S^5$ \cite{rev, rev2} is a reason for adopting the same approach for other less supersymmetric string backgrounds which are still integrable \cite{sym, sym2, sym3, sym4}. This approach has allowed tremendous progress in $AdS_4 \times \mathbb{CP}^3$ \cite{Klose:2010ki} and $AdS_3 \times S^3 \times {\cal{M}}^4$ \cite{AleS}. One of the most recent frontiers seems to be the Type II $AdS_2 \times S^2 \times T^6$ background with R-R fluxes, preserving 8 supersymmetries \cite{ads2, ads22, ads23, ads24, ads25}. The holographic dual is either expected to be a superconformal quantum mechanics, or a chiral 2D CFT \cite{dual, dual2, dual3, dual4} (see \cite{gen, gen2, gen3, gen4, gen5, gen6, gen7, gen8, gen9, gen10, gen11, gen12} for recent work). 

The $AdS_2 \times S^2$ (coset) part of the background is described by a
Metsaev-Tseytlin \cite{Metsaev:1998it} type of action \cite{sc, sc2} based on the quotient supergroup
$$
\frac{PSU(1,1|2)}{ SO(1,1) \times SO(2)}.
$$
The algebra $\mathfrak{psu}(1,1|2)$ admits a $\mathbb{Z}_4$ automorphism (cf. \cite{Bena:2003wd,Babichenko:2009dk}). This is crucial for classical integrability, which has been explicitly shown up to quadratic order in the fermions \cite{Sorokin:2011rr}, cf. also \cite{Cagnazzo:2011at, Cagnazzo:2011at2}.

In \cite{Hoare:2014kma}, an exact S-matrix was constructed for the scattering of excitations transforming under the centrally-extended $\mathfrak{psu}(1|1)^2$ symmetry superalgebra preserving the BMN vacuum
\cite{Berenstein:2002jq,amsw, amsw2}, generalising to $AdS_2/CFT_1$ the armamentarium familiar from higher dimensions \cite{beis0,conv,tor,tor2,Janik:2006dc,Borsato:2013qpa,Ya3,mf, mf2, mf3, mf4}. The S-matrix for massive modes is subject to crossing and unitarity, but a solution for the dressing factor is still unknown. Under certain assumptions, the near-BMN expansion gives consistency with perturbation theory \cite{amsw, amsw2}.

The principal feature which differentiates $AdS_2$ from its higher-dimensional relatives $AdS_5$, ($AdS_4$) and $AdS_3$, is that the magnon representations are {\it long} for any non-zero mass. Furthermore, because of generic reducibility of the tensor-product representation, the S-matrix depends on an undetermined function, which is fixed by the Yang-Baxter equation (cf. also \cite{Arutyunov:2009pw,pr, pr2}). The massless representations are instead {\it short}, and the approach adopted in \cite{Hoare:2014kma} - cf. \cite{BogdanLatest, BogdanLatest2, BogdanLatest3, BogdanLatest4} - is to take a {\it zero-mass / finite $h$} limit of the corresponding massive centrally-extended $\mathfrak{psu}(1|1)$ building block (cf. \cite{Zamol2, Zamol22}). One obtains in this fashion the limiting S-matrices for all choices of left- and right- movers. The scattering theory enjoys Yangian symmetry - cf. also \cite{Hoare:2015kla}.

\subsection{\label{sec:IB}Relativistic massless scattering}
Integrable massless scattering is a remarkable chapter in the theory of exact S-matrices and solvable two-dimensional models. It was pioneered by Zamolodchikov as a tool to describe the renormalisation group flow between conformal field theories, along massless integrable trajectories \cite{Zamol2, Zamol22}. The properties of massless S-matrices are slightly different from the standard ones valid for massive theories, and are summarised in \cite{Borsato:2016xns}, app.A. For the gauge-fixed string sigma-model, the situation is complicated by the absence of relativistic invariance. On the one hand, this allows for a perturbative description of scattering among only left or right movers \cite{BogdanLatest, BogdanLatest2, BogdanLatest3, BogdanLatest4}. On the other hand, the formulas, especially for $AdS_2$, are considerably more complicated.

To gain some insight in a simplified setting, we study here first the relativistic ({\it i.e.} strict BMN) limit of the massless $AdS_2$ S-matrix with ${\cal{N}} = 1$ supersymmetry, connected to ${\cal{N}}=1$ supersymmetric theories in $1+1$ dimensions. Massive representations and their scattering have been studied in \cite{Fendley:1990cy}. We will show how the S-matrices obtained in the massless relativistic limit from $AdS_2$ are related to those of \cite{Fendley:1990cy}. In the process, we will understand how to interpret their features in the light of the standard theory of massless scattering. This in turn clarifies some facts observed in the non-relativistic case.

\subsection{\label{sec:IC}Problems with the reference state}
The difficulties associated to $AdS_2$ do not terminate with the complicated form of the scattering matrix, which is attenuated by taking the massless limit. The specific form of the R-matrix\footnote{Throughout the paper, we will use both terminologies of ``R-matrices" and ``S-matrices". They contain the same physical information. The precise relationship for our purposes is given in section \ref{sec:VIA}.} closely resembles, in the location of its non-zero entries, the R-matrix for the XYZ spin-chain or the eight-vertex model \cite{Baxter:1972hz, Baxter:1972hz2}. This is a common feature to integrable models with ${\cal{N}}=1$ supersymmetry \cite{Schoutens,MC}. It is then easy to see that such R-matrices, and the associated transfer matrices, suffer from the problem of not admitting a reference state ({\it pseudo-vacuum}), namely, a lowest-weight state from which to construct the spectrum of excitations by means of the algebraic Bethe ansatz \cite{Levkovich-Maslyuk:2016kfv}. As a result, a set of Bethe equations, to match with those conjectured from the sigma-model in \cite{Sorokin:2011rr}, has not yet been constructed from the R-matrix.  

A variety of approaches have been developed in the literature to overcome such problems, starting with Baxter's original method of functional relations. Recently, \cite{Nepo, Nepo2, Nepo3} successfully applied this strategy to non-diagonal boundary problems, which suffer from the same issue (see also {\it e.g.} \cite{Belliard} for one particular example out of the vast related literature). Unfortunately, we do not yet have the appropriate Baxter operators for $AdS_2$. In \cite{Takhtajan:1979iv}, Faddeev and Takhtajan found a way of extending the algebraic Bethe ansatz to treat the specific case of the XYZ chain. A similar idea was used in the supersymmetric case, for those particular S-matrices which satisfy the so-called {\it free-fermion condition} \cite{Bazhanov:1984iw,Felderhof, Felderhof2, Felderhof3}. This condition allows a basis-transformation, convenient to  ultimately extract the eigenvalues the transfer matrix. Such scheme has been performed for ${\cal{N}}=1$ models in \cite{MC,Ahn:1993qa}, which we will closely follow. What this will produce for us is a set of (auxiliary) Bethe-ansatz conditions, identifying the location of potential zeroes of the transfer-matrix eigenvalues. We will manage to match this with a naive massless limit of the Bethe ansatz conjectured in \cite{Sorokin:2011rr}. The whole problem is then reduced to solving a factorisation condition ({\it inversion relation}, cf. \cite{Zamolodchikov:1991vh}), which we simply state. 

The free-fermion condition holds for the $AdS_2$ {\it massive} S-matrix as well, and it is nothing else than the determinantal condition which was observed in \cite{Hoare:2014kma}. Thanks to this condition, the S-matrix enjoys an accidental $\mathfrak{u}(1)$ symmetry under which only the fermions are charged, corresponding to a property of the string theory \cite{amsw, amsw2,Hoare:2014kma}. This in fact allows for the existence of a reference state for the {\it complete} scattering matrix, which consists of two copies of the one we shall study in this paper. Such phenomenon is echoed in other integrable systems \cite{Faddeev:1995nf} - see \cite{Davide} for a comprehensive discussion -, where a reference state is for instance found considering two or more adjacent spin-chain sites\footnote{We thank Davide Fioravanti for communication about this point.}. It is however here technically rather complicated to proceed that way\footnote{We thank Ben Hoare, Marius de Leeuw and Alessandro Sfondrini for communication about this point.}. Nevertheless, the free-fermion condition allows us to make progress in the massless case using the strategy of \cite{MC,Ahn:1993qa}. There is all hope that it will likewise help in the massive situation, although its implementation for long representations looks challenging at the moment. 

\subsection{\label{sec:ID}This paper}
Here is the outline of the paper. In section \ref{sec:II}, we recollect salient facts about the centrally-extended $\mathfrak{psu}(1|1)$ algebra governing the scattering of $AdS_2$ magnons, and specialise it to its massless limit. In section \ref{sec:III}, we perform the relativistic (BMN) limit, and display a family of relativistic massless S-matrices. Some of them are known from the theory of ${\cal{N}}=1$ supersymmetry, while others are related to them in a special way, which we describe and comment upon from the viewpoint of the relativistic theory of massless integrable scattering. In section \ref{sec:IIB}, we obtain these special S-matrices as the relativistic limit of those derived in \cite{Hoare:2014kma} for massless non-relativistic $AdS_2$ superstrings. In particular, we clarify a series of observations made in \cite{Hoare:2014kma} regarding the braiding-unitarity properties of these massless S-matrices, by analysing the issue from the relativistic viewpoint. In section \ref{sec:IV}, we write down the relativistic crossing symmetry condition, and find a minimal solution for the dressing factor, which we expect to be connected to the relativistic limit of the (yet unknown) dressing factor for left-left and right-right $AdS_2$ massless magnons. In section \ref{sec:V}, we obtain a set of differential equations for various relativistic S-matrices, in preparation to a similar type of geometric analysis as the one contained in \cite{Joakim,Andrea,Charles, Charles2,Riccardo}, which we plan for future work. In section \ref{sec:VI}, we illustrate the Bethe-ansatz technique based on the free-fermion condition, first displaying it for a known ${\cal{N}}=1$ S-matrix in section \ref{sec:VIA}, then specialising the process to the S-matrices obtained from $AdS_2$ in the BMN limit in section \ref{sec:VIB}. In section \ref{sec:VIC}, we generalise the procedure to the massless {\it non-relativistic} $AdS_2$ case, which is seen to work in a similar fashion. We obtain a particular set of Bethe-ansatz conditions, which we can compare with a naive massless limit of the one conjectured by Sorokin, Tseytlin, Wulff and Zarembo \cite{Sorokin:2011rr}. We then conclude with some remarks and open questions.

\section{\label{sec:II}Massless representation of centrally-extended $\mathfrak{psu}(1|1)$}
In this section we shall derive the massless representation of the centrally-extended $\mathfrak{psu}(1|1)$ symmetry algebra relevant for scattering of massless modes in $AdS_2\times S^2\times T^6$. We will first consider the massive representation, then apply the massless limit while remaining non-relativistic, and finally take the relativistic (BMN) limit. 

The centrally-extended $\mathfrak{psu}(1|1)$ Lie superalgebra is defined by the commutations relations:
\begin{eqnarray}
\{ \mathfrak{Q}, \mathfrak{Q} \} = 2\mathfrak{P}, \qquad 
\{ \mathfrak{S}, \mathfrak{S} \} = 2 \mathfrak{K}, \qquad
\{ \mathfrak{Q}, \mathfrak{S} \} = 2 \mathfrak{C},
\end{eqnarray}
where $\mathfrak{P}, \mathfrak{K}$ and $\mathfrak{C}$ are central bosonic generators, $\mathfrak{Q}$ and $\mathfrak{S}$ are fermionic generators. We begin by remarking that, from a purely algebraic viewpoint, these commutation relations already identify ${\cal{N}}=1$ supersymmetry in 1+1 dimensions \cite{Fendley:1990cy,Schoutens,MC} (see also \cite{Fok} and references therein), which we will take in various massless and relativistic representations in what follows. 

The (boson,fermion) $\big(|\phi\rangle, |\psi\rangle\big)$ representation we consider is the following:
\begin{align}
& \mathfrak Q\left|\phi\right>=a\left|\psi\right>, & & & &
\mathfrak Q\left|\psi\right>=b \left|\phi\right>, \nonumber
\\ 
&\mathfrak S\left|\phi\right>=c\left|\psi\right>, & & & &  
\mathfrak S\left|\psi\right>=d\left|\phi\right>, \nonumber
\\ 
&\mathfrak C\left|v\right>= C\left|v \right>, & & & &
\mathfrak P\left|v\right>=P \left|v \right>, & & & & 
\mathfrak K\left|v\right>= K \left|v \right>, 
\end{align}
where $v = \phi, \psi$, and $a, b, c, d, C, P$ and $K$ are the representation parameters:
\begin{align}
a =& \frac{\alpha \,e^{\frac{i\rm p}{4}-\frac{i\pi}4}}{\sqrt2} \sqrt{\rm e+\rm m},  &\nonumber
b =& \frac{\alpha^{-1} e^{-\frac{i\rm p}{4}+\frac{i\pi}4}}{\sqrt2} \frac{\rm h(1-e^{i\rm p})}{\sqrt{\rm e+\rm m}}, \\
c =\, & \frac{\alpha\, e^{\frac{i\rm p}{4}-\frac{i\pi}4}}{\sqrt2} \frac{\rm h(1-e^{-i\rm p})}{\sqrt{\rm e+\rm m}}, &
d =\, & \frac{\alpha^{-1}e^{-\frac{i\rm p}{4}+\frac{i\pi}4}}{\sqrt 2} \sqrt{\rm e+\rm m},\nonumber 
\end{align}
\begin{eqnarray}
\label{par2_massive}
C = \frac{\rm e}{2}, \qquad P = \frac{\rm h}{2} (1 - e^{i \rm p} ), \qquad K = \frac{\rm h}{2} (1 - e^{-i \rm p}), 
\end{eqnarray}
where $\rm e, \rm p$ and $m$ are the energy, spatial momentum and mass respectively, $\rm h$ is the coupling constant and $\alpha$ is an undetermined phase. One has the ``dispersion relation"
\begin{eqnarray}
\label{hence}
\rm e^2 = m^2 + 4 \rm h^2 \sin^2 \frac{\rm p}{2}. 
\end{eqnarray}
We recall that the $AdS_2$ massive representation we have just described  is {\it long}, hence (\ref{hence}) is not to be seen as a shortening condition. In particular, the ``mass" parameter $m$ is completely unconstrained. Its meaning as a mass has to be imposed by hand, inspired by analogy with the higher-dimensional cases.

The {\it non-relativistic massless} representation of centrally-extended $\mathfrak{psu}(1|1)$ is obtained by performing the limit $m \rightarrow 0$ at finite $h$. Setting $m = 0$ amounts to a shortening condition \cite{Hoare:2014kma}, which is expected to be protected from quantum corrections in the complete theory \cite{Sorokin:2011rr,Cagnazzo:2011at, Cagnazzo:2011at2}. The representation parameters become 
\begin{eqnarray}
\label{par1_massless}
a &=&\alpha \,e^{\frac{i\rm p}{4}-\frac{i\pi}4} \sqrt{\rm h \sin (\rm p / 2) },  \qquad\qquad b = \pm \frac{1}{\alpha} e^{\frac{i\rm p}{4} -\frac{i\pi}4} \sqrt{\rm h \sin (\rm p / 2)}, \nonumber \\
c &=& \pm \alpha\, e^{-\frac{i\rm p}{4}+\frac{i\pi}4} \sqrt{\rm h \sin (\rm p / 2) }, \qquad\ \  d = \frac{1}{\alpha}e^{-\frac{i\rm p}{4}+\frac{i\pi}4} \sqrt{\rm h \sin (\rm p / 2)}, 
\end{eqnarray}
where the upper sign is for right movers, the lower sign for left movers (in the latter case one also needs to account for a global factor of $\sqrt{-1} = i$ according to our choice of branch, which will matter in the mixed right-left and left-right coproducts). The massless dispersion relation is 
\begin{eqnarray}
\rm e = 2 \rm h \, \Big\lvert\sin \frac{\rm p}{2}\Big\rvert,
\end{eqnarray} 
$\mathfrak{Re} p \in (0,\pi)$ for right movers, $\mathfrak{Re} p \in (-\pi,0)$ for left movers. Crossing is implemented by first shifting the fundamental domain to $\mathfrak{Re} p \in (0,2 \pi)$, and then analytically continuing to negative momenta and energies, starting from the fundamental domain and going through the lower imaginary axis \cite{Borsato:2016xns}. As in $AdS_5$, the magnon representation is self-conjugate, while in $AdS_3$ crossing symmetry exchanges $L$ and $R$ particles in addition to analytically continuing the momenta.

\subsection{\label{sec:III}Relativistic massless R-matrices}
In this section we derive R-matrices in the relativistic massless limit.  We rescale the momentum and the coupling as follows
\begin{eqnarray}
\label{rel_lim}
\rm p \rightarrow \varepsilon q, \qquad\qquad \rm h \rightarrow \frac{c}{\varepsilon}, \qquad\qquad \varepsilon\rightarrow 0^+. 
\end{eqnarray}
where 
\begin{eqnarray}
q= e^{\theta},
\end{eqnarray}
$\theta$ is the rapidity and $c$ is the speed of light. The {\it relativistic massless representation} is then given by
\begin{eqnarray}
\label{par1massless}
\nonumber
a =&& \alpha \,e^{-i\frac{\pi}{4}}\sqrt{\frac{c \, q}{2}}, 
\qquad b = \pm \alpha^{-1}\,e^{-i\frac{\pi}{4}}\sqrt{\frac{c \, q}{2}}, \\
c =&& \pm \alpha \,e^{i\frac{\pi}{4}}\sqrt{\frac{c \, q}{2}}, 
\qquad d = \alpha^{-1} \,e^{i\frac{\pi}{4}}\sqrt{\frac{c \, q}{2}},   
\end{eqnarray}
where the upper sign is for right movers, the lower sign for left movers, and 
\begin{eqnarray}
\label{par2_massless_rel}
C = \frac{c \, q}{2}, \quad\qquad P = -i\frac{c \, q}{2}, \quad\qquad K = i\frac{c \, q}{2}, 
\end{eqnarray}
with the relativistic massless dispersion relation
\begin{eqnarray}
e = c \, |q|, 
\end{eqnarray}
where $\mathfrak{Re} q>0$ for right movers, $\mathfrak{Re} q<0$ for left movers. Relativistic invariance will guarantee all the R-matrices to only depend on the difference of the rapidities of the two scattering particles. 

The action of the symmetry on two-particle states in the relativistic limit is 
\begin{eqnarray}
\label{rel_coproduc}
\notag
&&\Delta(\mathfrak{Q}) = \mathfrak{Q} \otimes \mathbf{1} + \mathbf{1} \otimes \mathfrak{Q}, \qquad \Delta (\mathfrak{S} ) = \mathfrak{S} \otimes \mathbf{1} + \mathbf{1} \otimes \mathfrak{S},  \\
&&
\Delta (\mathfrak{P}) = \mathfrak{P} \otimes \mathbf{1} + \mathbf{1} \otimes \mathfrak{P}, 
\qquad \Delta( \mathfrak{C}) = \mathfrak{C} \otimes \mathbf{1} + \mathbf{1} \otimes \mathfrak{C},\nonumber\\ 
&&\quad \qquad \qquad \qquad \Delta ( \mathfrak{K} ) = \mathfrak{K} \otimes \mathbf{1} + \mathbf{1} \otimes \mathfrak{K},
\end{eqnarray}
where all the non-trivial {\it braiding} factors \cite{Hoare:2014kma,tor, tor2} have gone to $1$ when $\epsilon \to 0$. We impose the R-matrix to commute with the algebra action on two-particle states, 
\begin{eqnarray}
\label{com_massles_rel}
\Delta^{op} ( \mathfrak{J} ) R = R \Delta (\mathfrak{J}) , 
\end{eqnarray}
which must be true for all generators $\mathfrak{J}$ of centrally-extended $\mathfrak{psu}(1|1)$. Conservation of the total fermionic number constrains the R-matrix to be parametrised as
\begin{eqnarray}
\notag
R &=& A_{12} \, \mathsf{E}_{11} \otimes \mathsf{E}_{11} 
+ B_{12} \, \mathsf{E}_{11} \otimes \mathsf{E}_{22}
+ C_{12} \, \mathsf{E}_{22} \otimes \mathsf{E}_{11}\\
&+& D_{12} \, \mathsf{E}_{22} \otimes \mathsf{E}_{22}
+ Y_{12} \, \mathsf{E}_{12} \otimes \mathsf{E}_{12}
+ F_{12} \, \mathsf{E}_{21} \otimes \mathsf{E}_{21}\nonumber \\
&+& G_{12} \, \mathsf{E}_{12} \otimes \mathsf{E}_{21}
+ H_{12} \, \mathsf{E}_{21} \otimes \mathsf{E}_{12},
\end{eqnarray}
with $\mathsf{E}_{ij}$ the $2 \times 2$ {\it matrix unities}, {\it i.e.} matrices with all zeroes but $1$ in position $(i,j)$. The subscript ${}_{12}$ in the coefficients refers to particles $1$ and $2$ in the scattering. Eq. (\ref{com_massles_rel}) then implies that, if we consider for instance right-right scattering, then
\begin{eqnarray}
\notag
A_{12} &=& D_{12} - \frac{1}{\alpha^2} e^{\frac{\theta}{2}} F_{12} - \alpha^2 e^{-\frac{\theta}{2}} Y_{12} - 2 \cosh\frac{\theta}{2} G_{12}, \\
\notag
B_{12} &=& D_{12} - \alpha^2 e^{-\frac{\theta}{2}} Y_{12} - e^{-\frac{\theta}{2}} G_{12}, \\
\notag
C_{12} &=& D_{12} - \frac{1}{\alpha^2} e^{\frac{\theta}{2}} F_{12} - e^{\frac{\theta}{2}} G_{12},  \\
H_{12} &=& - \alpha^2 Y_{12} - \frac{1}{\alpha^2} F_{12} - G_{12},
\end{eqnarray}
and similarly for the other combinations. Whenever it is unambiguous from the context, we will always denote $\theta \equiv \theta_1 - \theta_2$.

We provide in what follows a list of R-matrices which are invariant under the relativistic massless coproduct action (\ref{rel_coproduc}) for the values of $\alpha$ specified below, and which satisfy the Yang-Baxter equation. Let us remark that, throughout all of section \ref{sec:II}, whenever we write matrices, we mean the matrices of coefficients of the R-matrix action on states. 
All matrices except for Solution 1 (meaning, Solutions 2 - 5) are for both particles with the same chirality (right-right or left-left, as specified case by case).

The way we have obtained the solutions reported below is by first imposing the symmetry w.r.t. the centrally extended $\mathfrak{psu}(1|1)$ algebra, and afterwards imposing the Yang-Baxter equation (YBE). Only then we have tested crossing symmetry (more precisely, Zamolodchikov's combined cross-unitarity condition \cite{Borsato:2016xns}) and braiding-unitarity, the latter corresponding to the property
\begin{eqnarray}
R_{12}(p_1,p_2) R_{21}(p_2,p_1) = \mathbf{1}, \qquad R_{21}(p,q) = R^{op}(p,q). 
\end{eqnarray} 
We have not exhausted all possibilities of solution of the YBE, and retained only those which will be relevant for our discussion. All our solutions satisfy cross-unitarity, but some will not satisfy braiding unitarity. We remark that \cite{Schoutens} found the most general solution to the YBE, satisfying the ${\cal{N}}=1$ supersymmetry algebra. The S-matrices on which \cite{Schoutens} focuses, however, all satisfy the properties of crossing symmetry {\it and} braiding unitarity, although the latter is dubbed ``optional".

\begin{itemize}
\item {\bf Solution 1}:  for arbitrary values of $\alpha$,
\begin{eqnarray}
\label{formula}
\begin{pmatrix}
1 & 0 & 0 & 0\\
0 & 1 & 0 & 0 \\
0 & 0 & 1 & 0 \\
0 & 0 & 0 & 1
\end{pmatrix}.
\end{eqnarray}
It trivially satisfies crossing symmetry and braiding unitarity. We shall see that the mixed scattering matrix of $AdS_2$ massless modes equals (\ref{formula}) in the relativistic limit, which signals a decoupling of left and right movers. 

\item {\bf Solution 2 (``Fendley $p=\frac{1}{2}$")}: for $\alpha^2 = 1$ (right-right), $\alpha^2 = -1$ (left-left),
\begin{eqnarray}
\label{Fendley p=1/2}
\begin{pmatrix}
1 & 0 & 0 & -\frac{\sinh \frac{\theta}{2}}{\cosh \theta}\\
0 & -\tanh\theta &\frac{\cosh\frac{\theta}{2}}{\cosh\theta}  & 0 \\
0 &\frac{\cosh\frac{\theta}{2}}{\cosh\theta}  & \tanh\theta & 0 \\
-\frac{\sinh \frac{\theta}{2}}{\cosh \theta} & 0 & 0 & - 1
\end{pmatrix}.
\end{eqnarray}
It satisfies crossing symmetry and braiding-unitarity. This is one of the solutions found in \cite{Fendley:1990cy}, and the ``$p = \frac{1}{2}$" parameter in the name reflects the notations of that paper - $p$ being no momentum at all in this case. It was there obtained for massive particles, but same-chirality relativistic R-matrices formally coincide with massive ones - cf. the discussion in \cite{Borsato:2016xns}.

The need to set the parameter $\alpha^2$ to a specific value in this solution (and, later on, in Solution 4), although in principle $\alpha$ should remain an undetermined relative scale between bosons and fermions, is to match with the precise S-matrices of \cite{Fendley:1990cy}, where a definite choice has been made for such scale.

\item {\bf Solution 3}: for arbitrary values of $\alpha$,
\begin{eqnarray}
\begin{pmatrix}
\label{left-right}
1 & 0 & 0 & \mp\alpha^{-2} e^{-\frac{\theta}{2}}\\
0 & -1 & e^{-\frac{\theta}{2}} & 0 \\
0 & e^{-\frac{\theta}{2}} & 1 & 0 \\
\mp\alpha^2 e^{-\frac{\theta}{2}} & 0 & 0 & -1 
\end{pmatrix},
\end{eqnarray}
where the upper sign is for right-right, the lower sign for left-left. It satisfies {\it cross-unitarity} \cite{Borsato:2016xns} (cf. also \cite{Bernard:1991vq}), but does \emph{not} satisfy braiding-unitarity. 
\item {\bf Solution 4 (``Fendley $p= - \frac{3}{2}$")}: for $\alpha^2 = 1$ (right-right), $\alpha^2 = -1$ (left-left),
\begin{eqnarray}
\label{Fendley p=-3/2}
\begin{pmatrix}
1 & 0 & 0 &  \frac{\sinh \frac{3}{2}\theta}{\cosh \theta}\\
0 &\tanh\theta  &\frac{\cosh\frac{3}{2}\theta}{\cosh\theta}  & 0 \\
0 & \frac{\cosh\frac{3}{2}\theta}{\cosh\theta}& - \tanh\theta   & 0 \\
\frac{\sinh \frac{3}{2}\theta}{\cosh \theta} & 0 & 0 & - 1
\end{pmatrix}.
\end{eqnarray}
It satisfies crossing symmetry and braiding-unitarity. This is  the other solution found in \cite{Fendley:1990cy}.
\item {\bf Solution 5}: for arbitrary values of $\alpha$,
\begin{eqnarray}
\begin{pmatrix}
\label{right-left}
1 & 0 & 0 & \pm \alpha^{-2} e^{\frac{\theta}{2}}\\
0 & 1 & e^{\frac{\theta}{2}} & 0 \\
0 & e^{\frac{\theta}{2}} & -1 & 0 \\
\pm \alpha^2 e^{\frac{\theta}{2}} & 0 & 0 & -1 
\end{pmatrix},
\end{eqnarray}
where the upper sign is for right-right, the lower sign for left-left. It satisfies cross-unitarity, but does \emph{not} satisfy braiding-unitarity. 
\end{itemize}
We remark that Solution 3, for right-right (resp., left-left) and $\alpha^2 = 1$ (resp., $\alpha^2 = -1$), can be obtained from Solution 2 as an asymptotic expansion at $\theta \rightarrow + \infty$, and from Solution 4 as an asymptotic expansion at $\theta \rightarrow - \infty$. Solution 5, instead, for right-right (resp., left-left) and $\alpha^2 = 1$ (resp., $\alpha^2 = -1$), can be obtained from Solution 2 as an asymptotic expansion at $\theta \rightarrow - \infty$, and from Solution 4 as an asymptotic expansion at $\theta \rightarrow + \infty$.

\subsection{\label{sec:IIB}Connection with non-relativistic massless scattering}
\label{connections}
There is a natural connection between the R-matrices found above and those which are solutions of the non-relativistic massless algebra given in \cite{Hoare:2014kma}. We verify here that Solutions 1, 3 and 5 descend from the non-relativistic massless-massless R-matrices found in \cite{Hoare:2014kma}.
 
We shall first remind the notation used in \cite{Hoare:2014kma}. The massless Zhukovsky variables $(x_1^{\pm}, x_2^{\pm})$ for two particles entering the scattering process satisfy the relation
\begin{eqnarray}
x_i^+ = \frac{1}{x_i^-}, \qquad i=1,2.
\end{eqnarray}
In our conventions, 
\begin{eqnarray} 
x_i^+ = \pm e^{i \frac{p_i}{2}},
\end{eqnarray}
where the upper sign stands for right movers $\mathfrak{Re} p_i \in (0, \pi)$, the lower sign for left movers  $\mathfrak{Re} p_i \in (-\pi,0)$. In \cite{Hoare:2014kma}, a function $f$ is introduced as
\begin{eqnarray}
f = \frac{\sqrt{\frac{x_1^+}{x_1^-}} \Big(x_1^- - \frac{1}{x_1^+}\Big) - \sqrt{\frac{x_2^+}{x_2^-}} \Big(x_2^- - \frac{1}{x_2^+}\Big)}{1 - \frac{1}{x_1^+ x_1^- x_2^+ x_2^-}}, 
\end{eqnarray}
controlling the value of the massive S-matrix entries. The function $f$ is not well-defined if we take the massless limit on both scattering particles, since $f$ assumes the indeterminate form $\frac{0}{0}$. The way one can compute $f$ in the massless-massless case is to perform the massless limit on one of the two particles and keep the other one massive, and finally perform the remaining massless limit. The order in which one performs the two limits matters \emph{only} for scattering of type right-right and left-left, while for scattering of type right-left and left-right such ambiguity does not appear. For the right-right and left-left cases, we will show that this mathematical ambiguity is connected to the two possibilities for Fendley's relativistic S-matrix \cite{Fendley:1990cy}. For the mixed cases, the ambiguity is absent and it guarantees that the BMN limit reproduces the trivial scattering matrix. We have verified that this pattern precisely matches the table given in section 5.2 of \cite{Hoare:2014kma}.

In order to show this, let us go one step backwards before the relativistic limit, and restore the non-triviality of the coproduct. On the supercharges (which generate all other coproducts by Lie superalgebra homomorphism) this reads \cite{Hoare:2014kma,tor, tor2}
\begin{eqnarray}
\label{rel_coproduc1}
&&\Delta(\mathfrak{Q}) = \mathfrak{Q} \otimes \mathbf{1} + e^{i \frac{p}{2}} \otimes \mathfrak{Q}, \quad \Delta (\mathfrak{S} ) = \mathfrak{S} \otimes \mathbf{1} + e^{-i \frac{p}{2}} \otimes \mathfrak{S}.  
\end{eqnarray}
The non-relativistic massless R-matrix in \cite{Hoare:2014kma} reads, in the mixed case,
\begin{eqnarray}
\label{Rnonrel}
R= \begin{pmatrix}
1 & 0 & 0 &  \pm \frac{1}{\alpha^2} \kappa(p_1,p_2)\\
0 & \pm \delta(x_1^+)  & \tilde{\kappa}(p_1,p_2) & 0 \\
0 & \tilde{\kappa}(p_1,p_2) & \mp \delta(x_2^{+}) & 0 \\
\pm \alpha^2\kappa(p_1,p_2) & 0 & 0 & - \delta(x_1^+) \delta(x_2^+)
\end{pmatrix},
\end{eqnarray}
where
\begin{eqnarray}
\delta(x_i^+) = \begin{cases} +1, & \mbox{if } x_i^+\mbox{ is right-mover} \\ -1, & \mbox{if } x_i^+\mbox{ is left-mover} \end{cases},
\end{eqnarray}
and
\begin{eqnarray}
\notag
\kappa(p_1,p_2) &=& -i \sqrt[4]{\frac{x_1^{+ 2}}{x_2^{+2}}} \frac{x_2^+ \sqrt{i(x_1^- - x_1^+)} \sqrt{i(x_2^- - x_2^+)}}{1- x_1^+ x_2^+ \pm (x_1^+ - x_2^+)},  \\
 \tilde{\kappa}(p_1,p_2) &=& 	\delta(x_1^+) \delta(x_2^+) \kappa(p_1, p_2).
\end{eqnarray}
In the \emph{right-left} massless scattering, $f \rightarrow 1$, one takes the upper sign in (\ref{Rnonrel}), and we obtain from it the matrix
\begin{eqnarray}
\label{RLrel}
\begin{pmatrix}
1 & 0 & 0 & 0\\
0 & 1 & 0 &  0 \\
0 & 0 & 1 & 0 \\
0 & 0 & 0 & 1 
\end{pmatrix},
\end{eqnarray}
which is the first-order term appearing in perturbation theory \cite{amsw, amsw2}. This is in agreement with Zamolodchikov's picture - cf. the discussion in \cite{Borsato:2016xns}. 

In the \emph{left-right} scattering, $f \rightarrow - 1$ and one takes the lower sign in (\ref{Rnonrel}). Again, the relativistic limit produces (\ref{RLrel}), which is once again consistent with Zamolodchikov's picture. 

The non-relativistic \emph{right-right} R-matrix reads
\begin{eqnarray}
\label{Rnonrel1}
R= \begin{pmatrix}
1 & 0 & 0 &  \pm \frac{1}{\alpha^2} \Big[\frac{\tan \frac{p_1}{4}}{\tan \frac{p_2}{4}}\Big]^{\pm \frac{1}{2}}\\
0 & \pm 1  & \Big[\frac{\tan \frac{p_1}{4}}{\tan \frac{p_2}{4}}\Big]^{\pm \frac{1}{2}} & 0 \\
0 & \Big[\frac{\tan \frac{p_1}{4}}{\tan \frac{p_2}{4}}\Big]^{\pm \frac{1}{2}} & \mp 1 & 0 \\
\pm \alpha^2\Big[\frac{\tan \frac{p_1}{4}}{\tan \frac{p_2}{4}}\Big]^{\pm \frac{1}{2}} & 0 & 0 & -1
\end{pmatrix}.
\end{eqnarray}
In the right-right massless scattering $f \rightarrow \pm 1$, which is a reflection of the massless limit order ambiguity mentioned above. The upper sign in (\ref{Rnonrel1}) corresponds to $f \to 1$, the lower to $f \to -1$. The relativistic limit of the massless R-matrix in \cite{Hoare:2014kma} gives: 
\begin{eqnarray}
\label{rightright5}
\begin{pmatrix}
1 & 0 & 0 & \frac{1}{\alpha^2} e^{\frac{\theta}{2}}\\
0 & 1 & e^{\frac{\theta}{2}} & 0 \\
0 & e^{\frac{\theta}{2}} & -1 & 0 \\
\alpha^2 e^{\frac{\theta}{2}} & 0 & 0 & -1 
\end{pmatrix}, \qquad f \rightarrow + 1, 
\end{eqnarray}
which reproduces Solution 5 right-right, and 
\begin{eqnarray}
\begin{pmatrix}
1 & 0 & 0 & - \frac{1}{\alpha^2} e^{-\frac{\theta}{2}}\\
0 & -1 & e^{-\frac{\theta}{2}} & 0 \\
0 & e^{-\frac{\theta}{2}} & 1 & 0 \\
-\alpha^2 e^{-\frac{\theta}{2}} & 0 & 0 & -1 
\end{pmatrix}, \quad f \rightarrow - 1, 
\end{eqnarray}
which reproduces Solution 3 right-right. 

Finally, the non-relativistic \emph{left-left} scattering reads
\begin{eqnarray}
\label{Rnonrel2}
R= \begin{pmatrix}
1 & 0 & 0 &  \pm \frac{1}{\alpha^2} \Big[\frac{\tan \frac{p_1}{4}}{\tan \frac{p_2}{4}}\Big]^{\mp \frac{1}{2}}\\
0 & \mp 1  & \Big[\frac{\tan \frac{p_1}{4}}{\tan \frac{p_2}{4}}\Big]^{\mp \frac{1}{2}} & 0 \\
0 & \Big[\frac{\tan \frac{p_1}{4}}{\tan \frac{p_2}{4}}\Big]^{\mp \frac{1}{2}} & \pm 1 & 0 \\
\pm \alpha^2\Big[\frac{\tan \frac{p_1}{4}}{\tan \frac{p_2}{4}}\Big]^{\mp \frac{1}{2}} & 0 & 0 & -1
\end{pmatrix}.
\end{eqnarray}
The upper sign in (\ref{Rnonrel2}) corresponds to $f \to 1$, the lower to $f \to -1$.
In the relativistic limit we obtain
\begin{eqnarray}
\label{leftleft3}
\begin{pmatrix}
1 & 0 & 0 &  \frac{1}{\alpha^2}e^{-\frac{\theta}{2}}\\
0 & -1 & e^{-\frac{\theta}{2}} & 0 \\
0 & e^{-\frac{\theta}{2}} & 1 & 0 \\
 \alpha^2 e^{-\frac{\theta}{2}} & 0 & 0 & -1 
\end{pmatrix}, \qquad f \rightarrow + 1, 
\end{eqnarray}
which reproduces Solution 3 left-left, and 
\begin{eqnarray}
\begin{pmatrix}
1 & 0 & 0 & -\frac{1}{\alpha^2} e^{\frac{\theta}{2}}\\
0 & 1 & e^{\frac{\theta}{2}} & 0 \\
0 & e^{\frac{\theta}{2}} &-1 & 0 \\
-\alpha^2 e^{\frac{\theta}{2}} & 0 & 0 & -1 
\end{pmatrix}, \qquad f \rightarrow - 1, 
\end{eqnarray}
which reproduces Solution 5 left-left.

We remind that the non-triviality of the massless right-right and left-left BMN limit is a completely non-perturbative effect, in full consonance with Zamolodchikov's picture of massless scattering \cite{Zamol2, Zamol22,Borsato:2016xns}.

The following diagram gives a snapshot of the relevant connections between the various limits of the R-matrices:  
{\normalsize
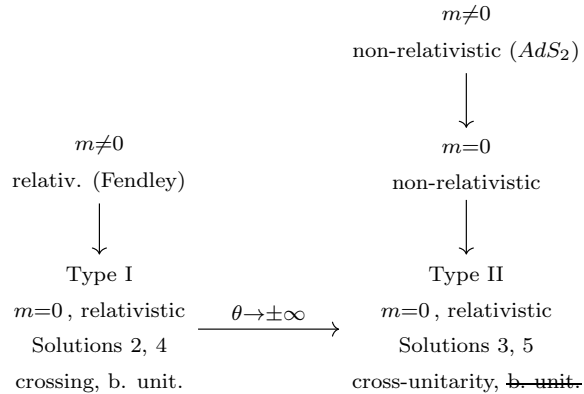
\begin{equation}
\begin{tikzcd}
 & & \substack{m \neq 0\\ \\ \text{non-relativistic ($AdS_2$)}}  \arrow{d}{} \\
 \substack{m \neq 0\\ \\ \text{relativ. (Fendley)}} \arrow{d}{}& &   \substack{m =  0\\ \\ \text{non-relativistic}} \arrow{d}{}\\
 \substack{\text{Type I} \\ \\ m = 0 \, ,\text{ relativistic}\\ \\ \text{Solutions 2, 4} \\ \\ \text{crossing, b. unit.}} \arrow{rr}{\theta \rightarrow \pm \infty} & &    \substack{\text{Type II} \\ \\ m = 0 \, ,\text{ relativistic}\\ \\ \text{Solutions 3, 5} \\ \\ \text{cross-unitarity, \st{b. unit.}}} 
\end{tikzcd} 
\nonumber
\end{equation}}

We recall that the BMN limit of the massive $AdS_2$ R-matrix is the identity. What we call {\it type I} relativistic massless R-matrices satisfy crossing and braiding unitarity, while the analogous {\it type II} R-matrices satisfy cross-unitarity, but do \emph{not} satisfy braiding unitarity. We have found that Solutions 3 and 5, which are of type II, can be obtained from Solutions 2 and 4, which are of type I, in the asymptotic expansion $\theta \rightarrow \pm \infty$, for appropriate values of $\alpha^2$. This implies that formally extracting the expression where $|\theta|$ is large will break braiding unitarity. Here we give an explanation, which clarifies a series of observations which were made in \cite{Hoare:2014kma}. 

Let $R^{(I)}$ be a R-matrix of type I, and let $R^{(II)}$ be a R-matrix of type II, such that, for example,
\begin{eqnarray}
R^{(I)} \sim \to R^{(II)} \quad \mbox{at $\theta \to \infty$}. 
\end{eqnarray}
Then $R^{(I)}$ must satisfy braiding unitarity, which is
\begin{eqnarray}
\label{buI}
R^{(I)}_{12} (\theta_1 - \theta_2) R^{(I)}_{21} (\theta_2 - \theta_1) = \mathbf{1}.
\end{eqnarray}
Now we take the asymptotic expansion $\theta \equiv \theta_1 - \theta_2 \rightarrow + \infty$ of both sides of (\ref{buI}). The first factor on the LHS of (\ref{buI}) asymptotes correctly to $R^{(II)}$ in the limit $\theta \rightarrow + \infty$. However, the second factor must be expanded at $- \infty$ in the limit $\theta\rightarrow + \infty$, which does not reproduce the solution $R^{(II)}$ as would be desired. It is also clear from this reasoning that, if we keep $f$ fixed, we have a form of combined braiding unitarity between Solution 3 and 5, which should physically correspond to the braiding unitarity of each of the two separate Fendley's solutions.

This is not inconsistent with Zamolodchikov's picture of massless scattering. In fact, if a massless relativistic S-matrix is obtained at large same-chirality (left-left and right-right) rapidities from a {\it massive relativistic} one, it shall be identical to the massive, hence will satisfy the same axioms. This applies to Fendley's solution, which is indeed identical to the massive theory \cite{Fendley:1990cy}. The other solutions obtained from $AdS$ are not subject to this argument, as they are obtained directly from non-relativistic ones, and coincide with the large-rapidity asymptotic expansion of \cite{Fendley:1990cy}.    

\begin{flushleft}
{\bf Comment} 
\end{flushleft}
There are two solutions which are invariant under the relativistic massless coproduct action (\ref{rel_coproduc}), which satisfy the Yang-Baxter equation, but which do not follow by any limit of the massless non-relativistic R-matrix in \cite{Hoare:2014kma}. These solutions are: 
\begin{itemize}
\item {\bf Solution 6}: for arbitrary values of $\alpha$ and $\kappa$ an arbitrary constant, 
\begin{eqnarray}
\begin{pmatrix}
1 & 0 & 0 & 0\\
0 & \frac{-1 + e^{\theta}}{-1 + e^{\theta} + e^{\kappa +\theta }} & \frac{e^{\kappa + \frac{\theta}{2}}}{-1 +e^{\theta} + e^{\kappa + \theta}} & 0 \\
0 & \frac{e^{\kappa + \frac{\theta}{2}}}{-1 + e^{\theta}+ e^{\kappa + \theta}} & 1 - \frac{e^{\kappa}}{-1 + e^{\theta} + e^{\kappa + \theta }} & 0 \\
0 & 0 & 0 & \frac{-1 - e^{\kappa} + e^{\theta}}{-1 + e^{\theta} + e^{\kappa + \theta}}
\end{pmatrix}.
\end{eqnarray}
It satisfies braiding-unitarity, but does \emph{not} satisfy crossing symmetry. 
\item {\bf Solution 7}: for $\alpha^2 = 1$,
\begin{eqnarray}
\begin{pmatrix}
1+  \frac{2i\sin \beta \pi  }{\sinh \theta} & 0 & 0 &  \frac{i\sin \beta \pi }{\cosh \frac{\theta}{2}}\\
0 & 1  & \frac{i \sin \beta \pi  }{\sinh \frac{\theta}{2}}  & 0 \\
0 & \frac{i \sin \beta \pi}{\sinh \frac{\theta}{2}}  & 1   & 0 \\
\frac{i \sin \beta \pi}{\cosh\frac{\theta}{2} } & 0 & 0 & 1 - \frac{2i \sin \beta \pi}{\sinh \theta}
\end{pmatrix}.
\end{eqnarray}
where $\beta$ is an arbitrary constant. It satisfies crossing symmetry and braiding-unitarity. This S-matrix corresponds to the supersymmetric Sinh-Gordon model \cite{SW,Ahn:1993qa,Ahn:1990uq,HRS} (with $\beta$ related to the coupling). We remark that the Pohlmeyer reduction of the $AdS_2$ superstring is the ${\cal{N}}=2$ supersymmetric sine-Gordon theory \cite{pr, pr2}, whose S-matrix is built from those of ${\cal{N}}=1$ supersymmetric sine-Gordon (Ben Hoare, private communication). It would be very interesting to explore this connection further in future work.
\end{itemize}

\section{\label{sec:IV}Crossing and Dressing Factors}
In this section, for definiteness, we will set $\alpha^2 = 1$ and focus on right-right scattering. Crossing symmetry is implemented in the following fashion. Define the supertranspose of a matrix $M$ as
\begin{equation}
M^{str}_{ij} = (-)^{ij+i} \, M_{ji},
\end{equation}
and the charge conjugation matrix as 
\begin{equation}
C = \mbox{diag}(i,1),
\end{equation}
such that 
\begin{eqnarray}
-\mathfrak{Q}_{q} = C^{-1} \mathfrak{Q}_{-q}^{str} \, C, \qquad -\mathfrak{S}_{q} = C^{-1} \mathfrak{S}_{-q}^{str} \, C,\nonumber \\
\end{eqnarray}
where the crossing map is given by
\begin{equation}
q \to - q, \qquad \theta \to i \pi + \theta.
\end{equation}

The cross-unitarity condition for Solution 2 reads
\begin{equation}
\label{cror}
R (\theta) \, \big[C^{-1}\otimes \mathfrak{1}\big] \, R^{str_1}(i \pi + \theta) \big[C \otimes \mathfrak{1}\big] = \bigg(1+\frac{\sinh^2 \frac{\theta}{2}}{\cosh^2 \theta}\Bigg) \mathfrak{1} \otimes \mathfrak{1} \ .
\end{equation}
From here, we deduce the cross-unitarity equation for the dressing factor $\Phi$:
\begin{equation}
\label{cror}
\Phi (\theta) \Phi(\theta + i \pi) = \bigg(1+\frac{\sinh^2 \frac{\theta}{2}}{\cosh^2 \theta}\Bigg) ^{-1} \ .
\end{equation}
The dressing factor given in \cite{Fendley:1990cy} is
\begin{eqnarray}
\label{Fe}
\Phi(\theta) &=& 4\bigg[\frac{1}{2} - \frac{\theta}{\pi i }\bigg]^2 \, \prod_{j=1}^\infty \frac{\Big(j-\frac{1}{2}\Big) \, \prod_{k=1}^{3} \Big(3j + \frac{1}{2} - k\Big)}{\Big(2j - \frac{1}{2}\Big)^2 \Big(2j + \frac{1}{2}\Big)^2 \Big(4 j^2 - \big[\frac{1}{2} - \frac{\theta}{\pi i }\big]^2\Big)^2}\nonumber\\
&&\times \frac{\Gamma\Big(3j-\frac{5}{2}+\frac{3}{2} \frac{\theta}{\pi i}\Big)\Gamma\Big(3j-1-\frac{3}{2} \frac{\theta}{\pi i}\Big)}{\Gamma\Big(3j-1+\frac{3}{2} \frac{\theta}{\pi i}\Big)\Gamma\Big(3j+\frac{1}{2}-\frac{3}{2} \frac{\theta}{\pi i}\Big)} \frac{\Gamma\Big(j-\frac{1}{2}+\frac{\theta}{2\pi i}\Big)\Gamma\Big(j-\frac{\theta}{2\pi i}\Big)}{\Gamma\Big(j+\frac{1}{2}-\frac{\theta}{2\pi i}\Big)\Gamma\Big(j+\frac{\theta}{2\pi i}\Big)} \ .
\end{eqnarray}
We have verified that (\ref{Fe}) solves equation (\ref{cror}). It also has the right analiticity structure, meaning no poles in the physical strip $\Im \theta \in (0,\pi)$ (as massless particles cannot form bound states).  

We have also constructed the minimal solution to the cross-unitarity equation satisfied by Solution 3, essentially by factorising Zamolodchikov's formula for the Sine-Gordon dressing factor \cite{Zamolodchikov:1977nu}. It can be directly verified, by using properties of the Gamma function and its product-representation, that the expression
\begin{eqnarray}
\Omega(\theta) = \frac{e^{\frac{\gamma}{2}- \frac{\pi i }{8}+\frac{\theta}{4}}}{\sqrt{2 \pi}} \prod_{j=1}^\infty e^{-\frac{1}{2 j}} \, j \, \frac{\Gamma\Big(j-\frac{1}{2}+\frac{\theta}{2\pi i}\Big)\Gamma\Big(j-\frac{\theta}{2\pi i}\Big)}{\Gamma\Big(j+\frac{1}{2}-\frac{\theta}{2\pi i}\Big)\Gamma\Big(j+\frac{\theta}{2\pi i}\Big)} \ , 
\end{eqnarray}
satisfies
\begin{equation}
\Omega (\theta) \Omega(\theta + i \pi) = \frac{e^{\frac{\theta}{2}}}{2 \cosh \frac{\theta}{2}} \ ,
\end{equation}
where $\gamma$ is Euler's constant. The factor $\Omega(\theta)$ has no poles in the physical strip. Hence, no CDDs are necessary, and we therefore expect it to be associated to the limit of the massless $AdS_2$ dressing factor in the corresponding relativistic limit, shadowing an analogous phenomenon occurring in the $AdS_3$ case \cite{Bogdan}.

\section{\label{sec:V}Differential equations}
The R-matrices (\ref{left-right}), (\ref{Fendley p=1/2}) and (\ref{right-left}), (\ref{Fendley p=-3/2}) respectively satisfy the following partial differential equations (for $\alpha^2=1$ and right-right scattering): 
\begin{eqnarray}
\label{PDE1}
\left[\frac{\partial}{\partial \theta} + \Gamma^{(i)}_{\theta}\right] R = 0 \ ,
\end{eqnarray} 
where
\begin{eqnarray}
\Gamma^{(3,5)}_{\theta} = \pm \frac{1}{2(1 + e^{\pm \theta})} \mathbf{1} \otimes \mathbf{1} - \frac{1}{4 \cosh(\frac{\theta}{2})} \sigma_1 \otimes \sigma_1\ ,
\end{eqnarray}
(with $3 / 5$ associated to the upper/ lower sign),
\begin{eqnarray}
\notag
\Gamma^{(2)}_{\theta} &=& \frac{\tanh\frac{\theta}{2}(2 - \cosh\theta)}{2\cosh\theta(1 - 2 \cosh\theta)} \mathbf{1}\otimes \mathbf{1} +\frac{2 - \cosh\theta}{2\cosh\frac{3\theta}{2}} (\mathsf{E}_{12} \otimes \mathsf{E}_{12} + \mathsf{E}_{21}\otimes\mathsf{E}_{21} )\\ 
&+&\frac{\cosh\frac{\theta}{2}}{1-2\cosh\theta} (\mathsf{E}_{12}\otimes\mathsf{E}_{21} + \mathsf{E}_{21}\otimes\mathsf{E}_{12} )\ , 
\end{eqnarray}
and 
\begin{eqnarray}
\notag
\Gamma^{(4)}_{\theta} &=& \bigg( \frac{\sinh\theta - 2 \sinh 2\theta}{1-2\cosh\theta + 2 \cosh 2\theta} - \frac{1}{2} \tanh\frac{\theta}{2} + \tanh\theta \bigg) ( \mathsf{E}_{11}  \otimes\mathsf{E}_{11} + \mathsf{E}_{22} \otimes \mathsf{E}_{22}) \\
\notag
&&- \frac{4\sinh\theta + \sinh 3\theta + 2\tanh\theta}{2(\cosh 2\theta + \cosh 3\theta)} (\mathsf{E}_{11} \otimes \mathsf{E}_{22} + \mathsf{E}_{22} \otimes \mathsf{E}_{11} ) \\
\notag
&& -  \frac{\cosh\frac{\theta}{2}(-3 + 2\cosh\theta)}{1- 2\cosh\theta + 2\cosh 2\theta} ( \mathsf{E}_{12} \otimes \mathsf{E}_{21} + \mathsf{E}_{21} \otimes \mathsf{E}_{12} ) \\
&&- \frac{3 - \cosh\theta + \cosh 2\theta}{2\cosh \frac{5\theta}{2}} (\mathsf{E}_{12}\otimes\mathsf{E}_{12} + \mathsf{E}_{21}\otimes \mathsf{E}_{21} ) \ ,
\end{eqnarray}
where $\sigma_1 \equiv \mathsf{E}_{12} + \mathsf{E}_{21}$. The (would-be) connections $\Gamma^{(i)}$ are meromorphic with poles in the complex $\theta$-plane. In the spirit of \cite{Joakim,Andrea}, we conjecture that the above represents the relativistic limit of a ``non-relativistic" fibre bundle, with a 2D torus as a base space - which is then decompactified and complexified to $\mathbb{C}^2 \ni (p_1,p_2)$ because of the analytic continuation of the dressing factor - and a $U(\mathfrak{su}(1 | 1))$ fibre\footnote{Where $U$ denotes the universal enveloping algebra.}. The hint for the existence of a non relativistic fibre bundle has been first found in \cite{Andrea} in the context of $AdS_3$. The algebraic tail in the equation (\ref{PDE1}) gives a hint for the existence of a relativistic fibre bundle in $AdS_2$ as well, which might come from a parental non relativistic fibre bundle via a relativistic limit. We conjecture that the relativistic limit might be responsible for the shrinking\footnote{We thank Jock McOrist and Martin Wolf for communication about this point.} of the base space $T^2$ to $S^1$, decompactified and complexified to $\mathbb{C} \ni \theta$. Our conjecture is supported by the fact that the R-matrices (\ref{Fendley p=1/2}), (\ref{left-right}), (\ref{Fendley p=-3/2}),  (\ref{right-left}) do not depend on the coordinate $\theta_1 + \theta_2$, but only on $\theta_1 - \theta_2$, which is a peculiar implication of the relativistic limit. Therefore the coordinate $\theta_1 + \theta_2$ is mute, the connection $\Gamma$ along $\theta_1 + \theta_2$ is identically zero, hence the part of the fibre bundle constructed over the coordinate $\theta_1 + \theta_2$ can be disregarded, since it is geometrically trivial.

We have checked that the connections for Solution 3 / 5 respectively coincide with those for Solution 2 / 4 in the asymptotic large-$\theta$ regime, as expected from the discussion in section \ref{sec:II}. 

A very important point is that we have derived the above differential equations disregarding dressing factors, such as the solution $\Omega(\theta)$ obtained in the previous section. Adding them corrects the differential equations by adding a term proportional to the identity:
\begin{eqnarray}
\left[\frac{\partial}{\partial \theta} + \Gamma^{(i)}_{\theta} - \frac{\partial}{\partial \theta} \log \Omega(\theta) \right] \Omega(\theta) R = 0.
\end{eqnarray}
This can be interpreted, in the light of the conjecture we have advanced, as a $\mathfrak{u}(1) \subset \mathfrak{su(1|1)}$ shift of $- \frac{\partial}{\partial \theta} \log \Omega(\theta) \,  \mathbf{1} \otimes \mathbf{1}$ to the connection $\Gamma_\theta^{(i)}$. Whether this term has an interpretation analogous to a gauge transformation on a principle bundle is left for future investigation.

\section{\label{sec:VI}Bethe ansatz}
In this section, we study the Bethe-ansatz for the massless sector of $AdS_2$ superstrings. We start from the strict relativistic limit to illustrate the procedure, and eventually generalise the formulas to the non-relativistic case. We begin with Fendley's S-matrix \cite{Fendley:1990cy} for illustration purposes\footnote{We strongly believe that the Bethe ansatz for Fendley's S-matrix is known to experts (we acknowledge private communication with Zoltan Bajnok and Mauro Moriconi on this point). Given that we have not been able to explicitly retrieve it from the literature, we have decided to independently re-derive it and to display the detail of the procedure in section \ref{sec:VIA}, if only for illustrative scope.}. We then move on to those associated to the BMN limit of $AdS_2$ superstrings, and finally we generalise the technique to massless non-relativistic $AdS_2$.

\subsection{\label{sec:VIA}Free-fermion condition and basis-change}
As mentioned in the Introduction, the problem of lack of a reference state prevents the direct applicability of the algebraic Bethe ansatz technique to obtain the Bethe equations. The method relying on the {\it free-fermion condition}, which we will describe below, allows one to circumvent this problem. 

One finds it convenient to first switch from the R-matrix to a version of the S-matrix, and, after that, to proceed by {\it ignoring any further fermionic sign}. Whenever it has been possible to compare, we have checked that this preserves the Yang-Baxter equation and captures (a subset of) the same spectrum of eigenvectors, which would be obtained by the R-matrix based algebraic Bethe ansatz\footnote{We thank Ben Hoare for discussions about this point.} \cite{MC,Ahn:1993qa}. One writes such S-matrix as  
\begin{eqnarray}
S = \begin{pmatrix}A&B\\C&D\end{pmatrix}\ ,
\end{eqnarray}
where the matrix displayed acts on the first ({\it auxiliary}) space, while the operators $A$, $B$, $C$ and $D$ act on the second ({\it quantum}) space as follows:
\begin{eqnarray}
&&A = a_+ \, \mathsf{E}_{11} \, + \, b_+ \, \mathsf{E}_{22} \ , \qquad B = d_+ \, \mathsf{E}_{12} \, + \, c_- \, \mathsf{E}_{21}\ , \nonumber \\
&&C = c_+ \, \mathsf{E}_{12} \, + \, d_- \, \mathsf{E}_{21}\ , \qquad D = b_- \, \mathsf{E}_{11} \, + \, a_- \, \mathsf{E}_{22}\ .  
\end{eqnarray}
The functions appearing in the above formula read
\begin{eqnarray}
a_+ &=& a_- = 1 \ , \qquad\qquad\  b_- = - b_+ = \tanh \theta\ , \nonumber\\
c_+ &=& c_- = \frac{\cosh \frac{\theta}{2}}{\cosh \theta}\ , \qquad d_+ = - d_- = \frac{\sinh \frac{\theta}{2}}{\cosh \theta}\ .
\end{eqnarray}
One momentarily suppresses the dressing factor, which can easily be reinstated at the very end. The entries satisfy the {\it free-fermion condition}:
\begin{eqnarray}
a_+ a_- + b_+ b_- = c_+ c_- + d_+ d_-\ .
\end{eqnarray}
The monodromy matrix $M$ and the associated transfer matrix $T$ are defined as
\begin{eqnarray}
M =  S_{01} (\theta - \theta_1) ... S_{0N} (\theta - \theta_N)\ , \qquad T = \rm tr_0 M\ , 
\end{eqnarray}
and they constitute the basis for $RTT$ quantisation \cite{Levkovich-Maslyuk:2016kfv}. As we have already remarked, the very first step of the algebraic Bethe ansatz is hindered here by the impossibility of finding a natural lowest-weight eigenstate of $T$, from which the spectrum can be spanned by repeated action of the off-diagonal entries of $M$.
 
The strategy of \cite{Felderhof, Felderhof2, Felderhof3} and \cite{MC,Ahn:1993qa} goes as follows. First, define a new S-matrix $S^{(1)}$, which takes the same form as $S$ but replacing the functions by
\begin{eqnarray}
&&a_\pm \to a^{(1)}_\pm = - b_\pm\ , \qquad\, b_\pm \to b^{(1)}_\pm = a_\pm\ , \nonumber\\
&&c_\pm \to c^{(1)}_\pm = c_\pm\ , \qquad\quad d_\pm \to d^{(1)}_\pm = - d_\pm\ .
\end{eqnarray}  
One can promptly notice that $S^{(1)}$ still satisfies the free-fermion condition. Then, one writes
\begin{eqnarray}
\label{traccia}
T \, T^{(1)} &=& \rm tr_0 \Big[ S_{01} (\theta - \theta_1) ... S_{0N} (\theta - \theta_N) \big] \rm tr_{0'} \Big[S^{(1)}_{0'1} (\theta - \theta_1) ... S^{(1)}_{0'N} (\theta - \theta_N)\Big] \nonumber \\
&=& \rm tr_{0 \otimes 0'} \prod_{i=1}^N S_{0i} (\theta - \theta_i) \otimes S^{(1)}_{0'i}(\theta - \theta_i)\ ,  
\end{eqnarray} 
where the tensor product is between the two auxiliary spaces $0$ and $0'$ pertaining to $T$ and $T^{(1)}$, respectively.

The trick is now to find a similarity transformation on the combined tensor-product object $S_{0i} (\theta - \theta_i) \otimes S^{(1)}_{0'i}(\theta - \theta_i)$, capable of putting it into an upper triangular form. Such a transformation is performed at each site, but it should not depend on the site-specific variables ({\it inhomogeneities}) $\theta_i$. Because of the difference-form, this means that the similarity matrix should be a constant, which is a non-trivial step which requires the free-fermion condition. It is only in this fashion that the similarity matrices will all cancel in expression (\ref{traccia}), and the task of taking the trace will become straightforward.

In fact, one can prove that the following matrix:
\begin{eqnarray}
\label{x}
X = \frac{1}{\sqrt{2}} \begin{pmatrix}0&1&1&0\\1&0&0&1\\1&0&0&-1\\0&1&-1&0\end{pmatrix} = X^{-1} 
\end{eqnarray}
is such that
\begin{eqnarray}
\label{suchthatx}
X S_{0i} \otimes S^{(1)}_{0'i} X^{-1} &=& X \begin{pmatrix}A A^{(1)}&A B^{(1)}&B A^{(1)}&B B^{(1)}\\A C^{(1)}&A D^{(1)}&B C^{(1)}&B D^{(1)}\\C A^{(1)}&C B^{(1)}&D A^{(1)}&D B^{(1)}\\C C^{(1)}&C D^{(1)}&D C^{(1)}&D D^{(1)}\end{pmatrix} X^{-1} = \begin{pmatrix} m_+ & * & * & *\\0 & n_+ & * & * \\ 0 & 0 & n_- & * \\ 0 & 0 & 0 & m_-\end{pmatrix},
\end{eqnarray}
with
\begin{eqnarray}
&&m_\pm = \frac{1}{2 \cosh^2 (\theta - \theta_i)} \Big[\pm \cosh  (\theta - \theta_i) + \cosh 2 (\theta - \theta_i)\Big] \mathfrak{1}\ , \nonumber \\
&&n_\pm = \frac{1}{2 \cosh^2 (\theta - \theta_i)} \Big[\pm \sinh  (\theta - \theta_i) + \sinh 2 (\theta - \theta_i)\Big] \sigma_3\ ,
\end{eqnarray}
having denoted $\mathfrak{1} = \mathsf{E}_{11} + \mathsf{E}_{22}$ and $\sigma_3 = \mathsf{E}_{11} - \mathsf{E}_{22}$. Since
$\rm tr_{0 \otimes 0'} = \rm tr_4$, we can immediately write 
\begin{eqnarray}
\label{traccia2}
T \, T^{(1)} &=& \prod_{i=1}^N m_+ (\theta - \theta_i) +  \prod_{i=1}^N m_- (\theta - \theta_i) + \prod_{i=1}^N n_+ (\theta - \theta_i) + \prod_{i=1}^N n_- (\theta - \theta_i) \ .  
\end{eqnarray}  
We now make use of a very particular relation between $S$ and $S^{(1)}$. One can check that
\begin{eqnarray}
\label{rela}
&&S_{0'i}^{(1)} = \tau \, \sigma_1 \, S_{0'i}(\theta - \theta_i + i \pi) \sigma_1^{-1} \, \tau^{-1}\ , \nonumber \\
&&\sigma_1 = \mathsf{E}_{12} + \mathsf{E}_{21}\ , \qquad \tau = \mathsf{E}_{11} + i \mathsf{E}_{22}\ ,
\end{eqnarray}
where this similarity transformation is performed in the auxiliary space $0'$.
This means that 
\begin{eqnarray}
\label{that}
T^{(1)} (\theta) = T(\theta + i \pi)\ , 
\end{eqnarray}
which turns (\ref{traccia2}) into a crossing-type equation referred to as {\it inversion relation} \cite{Zamolodchikov:1991vh}. This is consistent with the property that $m_+$ maps into $m_-$ and $n_+$ into $n_-$, under $\theta \to \theta + i \pi$. We have performed explicit checks, for small $N$, that (\ref{traccia2}) is correct.

The eigenvalues of  (\ref{traccia2}) will be given by the same expression, with $\sigma_3$ replaced by the fermionic number of the particular eigenstate. The final task is then to factorise such expressions into a product of two functions, namely $f(\theta)f(\theta + i \pi)$. As it is familiar from solving crossing-symmetry relations for S-matrices, this is a difficult problem in general, whose study relies on analyticity assumptions. Here, we shall content ourselves with deriving a condition which identifies potential zeroes of the transfer-matrix eigenvalues, a condition which will be seen to lead to auxiliary Bethe-ansatz type equations. First, one splits 
\begin{eqnarray}
\label{times}
\notag
&&T T^{(1)} = \bigg[\prod_{i=1}^N A(\theta - \theta_i) + F \prod_{i=1}^NB(\theta - \theta_i)\bigg]\\
&&\times \bigg[\prod_{i=1}^N C(\theta - \theta_i) + F \prod_{i=1}^ND(\theta - \theta_i)\bigg]\frac{1}{\prod_{i=1}^N 2\cosh^2(\theta - \theta_i)}\ ,
\end{eqnarray}
where $F=\pm$ is the fermionic degree of the particular state one considers, and
\begin{eqnarray}
A(\theta - \theta_i) = \frac{c_{0i}^+}{C_{0i}}\ , \qquad B(\theta - \theta_i) = \frac{s_{0i}^-}{C_{0i}}\ , \qquad D(\theta - \theta_i) = \frac{s_{0i}^+}{c_{0i}^+} C_{0i}\ ,
\end{eqnarray}
where $C_{0i}$ is a freedom of this rewriting, and we have defined
\begin{eqnarray}
&&c_{0i}^\pm = \pm \cosh (\theta - \theta_i)+ \cosh 2  (\theta - \theta_i)\ , \nonumber \\
&&s_{0i}^\pm = \pm \sinh  (\theta - \theta_i)+ \sinh 2  (\theta - \theta_i)\ ,
\end{eqnarray}
and used the fact that
\begin{eqnarray}
\label{thanks}
c_{0i}^+ \, c_{0i}^- \, = \, s_{0i}^+ \, s_{0i}^-\ .
\end{eqnarray}
The eigenvalue has $N$ potential poles, determined by the hyperbolic cosines at the denominator, and a certain number of potential zeroes, possibly coincident, depending on the particular state. The possible location of them will be determined below via a set of auxiliary Bethe-ansatz conditions. Moreover, the eigenvalues are periodic of period $2 \pi i$, as can be seen by the fact that, by shifting (\ref{traccia2}) of a further $+i \pi$ and using (\ref{that}), one gets
\begin{eqnarray}
T(\theta + i \pi ) T(\theta + 2 i \pi) = T(\theta) T(\theta + i \pi)\ , 
\end{eqnarray}
where we have also explicitly made use of the invariance of the r.h.s. of (\ref{traccia2}) under shift of $i \pi$.
Therefore, $T(\theta + 2 i \pi) = T(\theta)$, hence it can be analysed by studying the location of its poles and zeroes in the strip $\theta \in [-\pi, \pi)$. We recall that the dressing factor - say, $\Phi$ - does not affect these considerations, since we already know how we will have to decorate the eigenvalue $T$ obtained at the end, {\it i.e.} by a product of dressing factors $\prod_{i=1}^N \Phi(\theta-\theta_i)$. From (\ref{times}), one sees that potential zeroes can come from
\begin{eqnarray}
\prod_{i=1}^N \frac{A(z_k - \theta_i)}{B(z_k - \theta_i)} = - F\ , \qquad \prod_{i=1}^N \frac{C(z_k - \theta_i)}{D(z_k - \theta_i)}= - F\ .
\end{eqnarray}
Plugging the explicite formulas, we see that the freedom of $C_{0i}$ is indeed irrelevant, and we obtain that the potential zeroes can come from either of two conditions:
\begin{eqnarray}
\label{aux}
\prod_{i=1}^N \coth \frac{z^+_k - \theta_i}{2} = - F\ , \qquad \prod_{i=1}^N \coth \frac{3(z^-_k - \theta_i)}{2} = - F\ ,
\end{eqnarray}
for any fixed value of the fermionic number $F=0,1$ of the state under consideration. Notice that, thanks to (\ref{thanks}), each of two auxiliary Bethe equations maps into itself under $z_k^\pm \to z_k^\pm + i \pi$.

One needs at this point to identify the actual set of zeroes of $T$ {\it vs.} those of $T^{(1)}$ in order to extract the eigenvalues of $T$. This can then be used to write down the ({\it momentum-carrying}) Bethe equation: intuitively, if one considers $N+1$ excitations on a circle of length $L$ with periodic boundary conditions, interacting via the scattering matrix $S$, one is led to consider a quantisation condition of the following type \cite{MC,Ahn:1993qa}:
\begin{eqnarray}
\label{eige}
e^{i p_0 L} \, T(p_0 | p_1,...,p_N) |\psi\rangle = |\psi \rangle\ ,
\end{eqnarray} 
where $p_a = e^{\theta_a}$, $a=0,...N$, and
\begin{eqnarray}
\label{tracea}
T(p_0|p_1,...,p_N) = \mbox{tr}_0 M(p_0|p_1,...,p_N)
\end{eqnarray} 
is the transfer matrix, supplemented by the appropriate dressing factors. Revolving the particle $0$ around the circle involves scattering all the other ones in sequence, returning the same eigenstate $|\psi \rangle$ of the transfer matrix. Eq. (\ref{eige}) is then turned into an equation for the eigenvalues. These are parametrised by their potential poles and zeroes, a subset of which are obtained from (\ref{aux}). 

Some experimenting with small $N$ seems to reveal that this final step is not straightforward, and would require a separate analysis. This will be so for all the cases which we discuss in this paper. This observation  is clearly related to the fact that our eigenvalues may either tend to zero (as in this section) or have an essential singularity (as in the next section) at $\theta = \infty$, at odds with the situation in \cite{MC,Ahn:1993qa} - and \cite{Zamolodchikov:1991vh}, after going to a {\it reduced} transfer matrix with no essential singularities. This is easily evinced by studying the asymptotics of the corresponding S-matrices. Therefore, even the knowledge of the zeroes and poles of the meromorphic periodic function $T(\theta)$ would not allow us to completely reconstruct it, as one cannot eliminate the ambiguity of factors which are entire periodic functions of $\theta$ and depend on all $\theta_i$'s. 

\subsection{\label{sec:VIB}Bethe-ansatz condition for Solution 3}
Now that we have illustrated the procedure and setup the notation, it is a simple exercise to apply it to Solution 3, which of course still satisfies the free-fermion condition. The advantage of having performed the process on Fendley's S-matrix is manifest from the fact that Solution 3 can formally be obtained from it in a large $\theta$ asymptotic expansion. This does not mean that one can indiscriminately expand at large $\theta$ all the previous formulas, but in several cases it implies that similar algebraic manipulations will apply.

We can be concise on the intermediate steps, and write now
\begin{eqnarray}
&&a_+ = a_- = 1\ , \qquad\quad b_- = - b_+ = 1\ , \nonumber\\
&&c_+ = c_- = e^{-\frac{\theta}{2}}\ , \qquad d_+ = - d_- = e^{-\frac{\theta}{2}}\ ,
\end{eqnarray}
such that
\begin{eqnarray}
a_+ a_- + b_+ b_- = c_+ c_- + d_+ d_-\ .
\end{eqnarray}
and
\begin{eqnarray}
&&a_\pm \to a^{(1)}_\pm = - b_\pm\ , \qquad b_\pm \to b^{(1)}_\pm = a_\pm\ , \nonumber\\
&&c_\pm \to c^{(1)}_\pm = c_\pm\ , \qquad\quad d_\pm \to d^{(1)}_\pm = - d_\pm\ .
\end{eqnarray}    
The very same transformation $X$ in (\ref{x}) and (\ref{suchthatx}) works for this case as well, and one obtains an upper triangular form like (\ref{suchthatx}) for $T T^{(1)}$, this time with diagonal entries
\begin{eqnarray}
&&m_\pm = \big(1\pm e^{-\theta}\big) \mathfrak{1}\ , \qquad n_\pm =  \big(1\pm e^{-\theta}\big) \sigma_3\ .
\end{eqnarray}
The great simplification with respect to the previous section is now that the product (\ref{traccia2}) of the two eigenvalues reduces to
\begin{eqnarray}
\label{fromm}
T T^{(1)} = (1+F) \Bigg[\prod_{i=1}^N \Big(1 + e^{-(\theta - \theta_i)}\Big) +   \prod_{i=1}^N \Big(1 - e^{-(\theta - \theta_i)}\Big)\Bigg]\ ,
\end{eqnarray}
where $F$ denotes again the fermionic number of the particular eigenstate under consideration. Formula (\ref{fromm}) shows that, in this case, we access only part of the spectrum, as $T T^{(1)}$ annihilates all fermionic eigenstates. The auxiliary Bethe equations, for each subset of $M$ $\beta_k$'s chosen amongst the potential zeroes of the bosonic transfer-matrix eigenvalues, read:
\begin{equation}
\label{sub}
\prod_{i=1}^N \tanh \frac{\beta_k - \theta_i}{2} = -1\ , \quad k=1,...,M\ .
\end{equation}
Furthermore, one can verify that the relations (\ref{rela}) and (\ref{that}) work exactly the same way, hence one can rely on the very same $2 \pi i$ periodicity property of the eigenvalue of $T$. One would then write the momentum-carrying equation
\begin{eqnarray}
\label{fixo}
e^{i e^{\theta_0} L} \, \Bigg[\prod_{i=1}^N \Omega(\theta_0 - \theta_i)\Bigg] \Lambda(\theta_0|\theta_1,...,\theta_N| \beta_1,...,\beta_M)= 1\ ,
\end{eqnarray}
subject to (\ref{sub}), where $\Lambda$ is the transfer-matrix eigenvalue normalised to $a_+=1$, and we have inserted the dressing factor obtained in section \ref{sec:IV}. As we shall see next, the condition (\ref{sub}) on the potential zeroes of $\Lambda$ matches the naive massless relativistic limit of the auxiliary Bethe equations for $AdS_2$ \cite{Sorokin:2011rr}.

\subsection{\label{sec:VIC}Bethe-ansatz condition for non relativistic massless $AdS_2$}
Let us now go back to the {\it non-relativistic} massless S-matrix. It is easy to see that we can simply repeat the entire line of argument of the previous section, with the replacement (consistent with the relativistic limit)
\begin{eqnarray}
e^{\frac{\theta - \theta_i}{2}} \to \sqrt{\frac{\tan \frac{p_0}{4}}{\tan \frac{p_i}{4}}}\ , \qquad\quad 1 \pm e^{\theta - \theta_i}\to 1 \pm \frac{\tan \frac{p_0}{4}}{\tan \frac{p_i}{4}}\ ,
\end{eqnarray} 
where $\theta \to \theta + i \pi$ is replaced by $p_0\to -p_0$, and under the square-root this is prescribed to give $i$. This means that the auxiliary Bethe equations now read
\begin{equation}
\label{just}
\prod_{i=1}^N \frac{\sin \frac{q_k + p_i}{4}}{\sin \frac{q_k - p_i}{4}} = - 1\ , \qquad k=1,...,M\ .
\end{equation}
If we naively take the massless relativistic limit of half (corresponding to one wing of the $\mathfrak{psu}(1,1|2)$ Dynkin diagram) of the auxiliary Bethe equations conjectured in \cite{Sorokin:2011rr} (STWZ), we find that they seem to exactly match with the {\it square} of (\ref{just}), if we identify their auxiliary roots, say, $p_{k,3}$ with our $q_k$, switching {\it e.g.} the type $1$ roots off. 

The momentum-carrying equation can easily be obtained via the same naive limit from STWZ, and it can be simplified using momentum conservation $\sum_{j=1}^N p_{j,2}=0$ (in their conventions, translating into $\sum_{j=1}^N p_{j}=0$ in ours) to be expressed in terms of the same functions appearing in (\ref{just}), except for the dressing factors. However, once again a separate analysis would be required to obtain such an equation from the procedure we have described, in analogy to (\ref{fixo}). The relativistic limit should then ideally reproduce some variant of (\ref{fixo}), precisely like (\ref{just}) straightforwardly reduces to (\ref{sub}). The proposal for the dressing factor to appear in the Bethe ansatz made in \cite{Sorokin:2011rr} involves the inverse-square of the BES factor \cite{BES, BES2}, which should then be compared with our section-\ref{sec:IV} $\Omega(\theta)$ in the appropriate massless relativistic limit along the lines of \cite{Bogdan}, something we have not yet attempted. 

\section{Conclusions}
In this paper, we have first analysed the relativistic scattering theory of massless excitations of the $AdS_2 \times S^2 \times T^6$ superstring \cite{Hoare:2014kma}, with the idea that this provides both a simplified setting where to resolve the complication of the system, and a warmup for attempting the derivation of the Bethe ansatz in the non-relativistic case. Using standard techniques available in the literature, one is capable of overcoming the issue of lack of a reference state. The so-called {\it free-fermion condition} is crucial in employing this strategy, and it appears to lie at the heart of the model, being a property of the massive theory as well associated to a $\mathfrak{u}(1)$ symmetry of the string theory \cite{Hoare:2014kma,amsw, amsw2}. Turning the argument around, we see that the partial embedding of ${\cal{N}}=1$ integrable S-matrices into string-theory might shed light on the nature of this rather miraculous constraint.

After having obtained a Bethe-ansatz condition for the relativistic case, we then discovered that it is almost straighforward to apply the same procedure to the massless non-relativistic situation. We could then compare with a naive massless limit applied to the conjecture of Sorokin, Tseytlin, Wulff and Zarembo \cite{Sorokin:2011rr}.

There are a series of natural open questions at this stage. The most urgent one is to solve the factorisation condition based on analiticity assumptions, and derive the complete set of Bethe equations for the massless {\it and} massive scattering theory, in such a way to perform (at least in the massive case) a thorough comparison with the Bethe equations of \cite{Sorokin:2011rr}. The free-fermion condition promises to be the essential tool to make progress, although the complication of the S-matrix entries will require a separate treatment. In particular, we expect it to be quite challenging to disentangle the eigenvalues of the transfer matrix $T$ from those of $T^{(1)}$ at the end of the process - cf. section \ref{sec:VI}. We plan this for future work.

Another striking feature of ${\cal{N}}=1$ (massive relativistic) supersymmetric models is Melzer's {\it folding} \cite{Melzer, Melzer2, Melzer3,MC}, which in fact relates the Thermodynamic Bethe Ansatz of ${\cal{N}}=2$ theories to the one of ${\cal{N}}=1$ theories. A similar ``folding" relationship exists between the dressing factors of the respective S-matrices. We are tempted to speculate a similar connection between $AdS_2$ and $AdS_3$ models  \cite{Bogdan}. We plan to eplore this avenue, and what it might entail for the respective string sigma models and their holographic duals, in upcoming work.

Finally, the analysis of the massless sector is of crucial importance to achieve a complete non-perturbative description of the theory. We believe that the non-triviality of the BMN limit in same-chirality scattering should play a key role in any progress in this direction \cite{Bogdan}.

Although difficult at the moment, it would be extremely interesting to find a connection between our setup and the recent developments concernig the SYK (Sachdev-Ye-Kitaev) model, where also an $AdS_2/CFT_1$ holographic problem is being attacked (see {\it e.g.} \cite{Maldacena:2016hyu}).

\section{Acknowledgments}
We would like to very much thank Bogdan Stefa\'nski, for crucial insights and illuminating discussions, Ben Hoare, for extremely useful discussions and detailed feedback on the manuscript, and Alessandro Sfondrini, for extremely valuable insights and encouragement. We thank Michael Abbott, Ines Aniceto, Samuel Belliard, Francesco Benini, Diego Bombardelli, Marius de Leeuw, Nick Dorey, Patrick Dorey, Davide Fioravanti, Daniel Grumiller, Romuald Janik, Jock McOrist, Sameer Murthy, Rafael Nepomechie, Olof Ohlsson Sax, Andrea Prinsloo, Vidas Regelskis, Martin Wolf for illuminating discussions. We thank Kostya Zarembo for discussions and extremely interesting remarks about the lack of a reference state in related problems. We thank Zoltan Bajnok and Mauro Moriconi for a very useful email exchange. We thank Joakim Str\"omwall for discussions and for help with the diagram of section \ref{connections}. 
A.F. is partially supported by the EPSRC grant FP/M506655. A.T. thanks the STFC under the Consolidated Grant project nr. ST/L000490/1. A.T. also thanks the Galileo Galilei Institute for Theoretical Physics (GGI) for the hospitality and INFN for partial support during the completion of this work, and everyone involved in the program {\it New Developments in $AdS_3/CFT_2$ Holography}, for stimulating discussions and a unique atmosphere.

\vskip 0.5cm
\noindent{\bf Data Management:} \vskip 0.1cm
\noindent  No data beyond those presented and cited in this work are needed to validate this study.
\vskip 0.5cm

\appendix

\end{document}